\documentclass[aps,prl,twocolumn,superscriptaddress]{revtex4-2}
\usepackage{amsmath}
\usepackage{graphicx}% Include figure files
\usepackage{dcolumn}% Align table columns on decimal point
\usepackage{bm}% bold math
\usepackage{xcolor}
\usepackage{hyperref}

\begin{document}

%\preprint{APS/123-QED}

\title{Cell-Dependent Criticality for Quantum Metrology}

%\title{Cell-Dependent Criticality for Quantum Metrology in Fock-Space Lattices}

%\title{Critical Quantum Metrology in a Fock-Space Lattice with Cell-Dependent Critical Response}

%\title{Critical Quantum Metrology from Cell-Dependent Critical Response in a Fock-Space Lattice}

%\title{Critical Quantum Metrology via Inhomogeneity-Organized Critical Response in a Fock-Space Lattice}

%\title{Critical Quantum Metrology via Distributed Critical Response in a Fock-Space Lattice}

%\title{Critical Quantum Metrology via a Trajectory-Response Mechanism in Fock Space}

\author{Zhoutao Lei}
\affiliation{Department of Physics, National University of Singapore, Singapore 117542, Singapore}

\author{Jihao Ma}
\affiliation{Laboratory of Quantum Engineering and Quantum Metrology, School of Physics and Astronomy, Sun Yat-Sen University (Zhuhai Campus), Zhuhai 519082, China}

\author{Yun Chen}
\affiliation{Laboratory of Quantum Engineering and Quantum Metrology, School of Physics and Astronomy, Sun Yat-Sen University (Zhuhai Campus), Zhuhai 519082, China}

\author{Tingting Wang}
\email{wangtingting@nnu.edu.cn}
\affiliation{Department of Physics, National University of Singapore, Singapore 117542, Singapore}
\affiliation{Phonon Engineering Research Center of Jiangsu Province, Ministry of Education Key Laboratory of NSLSCS, Center for Quantum Transport and Thermal Energy Science, Institute of Physics Frontiers and Interdisciplinary Sciences, School of Physics and Technology, Nanjing Normal University, Nanjing 210023, China}

\author{Jiangbin Gong}
\email{phygj@nus.edu.sg}
\affiliation{Department of Physics, National University of Singapore, Singapore 117542, Singapore}

\date{\today}% It is always \today, today,

\begin{abstract}
Exploiting enhanced sensitivity of a system in the vicinity of a phase transition boundary, critical quantum metrology to date still suffers from gap-closure related bottleneck effects, namely, critical slowing down of the sensing dynamics and a drastic shrinking of the parameter sensing window. To alleviate the said bottleneck inherent to any homogeneous lattice used for sensing, here we propose to leverage the intrinsic hopping inhomogeneity arising from bosonic ladder-operator matrix elements in Fock-space lattices (FSLs). Specifically, using a two-mode Jaynes--Cummings-type model, we show that the sensing parameter can be imprinted onto a topological zero-energy mode of the FSL. The key system parameters thus become cell dependent, effectively tracing out a curve in a topological phase diagram. Cell-dependent criticality emerges when this curve crosses or approaches a topological phase boundary, without globally tuning the lattice close to criticality. An external control parameter reshapes this curve, continuously tuning the scaling of the quantum Fisher information from the standard to the Heisenberg scaling while maintaining broad sensing coverage and a reduced gap cost. Furthermore, a local photon-number measurement on a single cavity saturates the quantum Fisher information. These results identify FSLs as a scalable and practical route to criticality-based quantum metrology. 
\end{abstract}

\maketitle 

{\it Introduction}.---Quantum metrology aims to improve the precision of estimating physical parameters relevant to precision sensing and fundamental tests~\cite{pezze2018quantum,braun2018quantum,montenegro2025quantum,ye2024essay}.  Entanglement-based metrology can in principle surpass the standard quantum limit (SQL)~\cite{wineland1992spin,kitagawa1993squeezed,lee2002quantum,giovannetti2004quantum,ma2011quantum,PhysRevA.110.022407}, but with highly entangled probes experimentally challenging to generate and protect~\cite{huelga1997improvement,kolodynski2010phase,escher2011general,chaves2013noisy,32n5-mhk1}. As a promising alternative, critical metrology harnesses the extreme sensitivity of physical systems near critical points to enhance estimation precision~\cite{campos2007quantum,zanardi2007information,zanardi2008quantum}.
Critical sensors have been explored across diverse forms of criticality, including first-order~\cite{raghunandan2018high,yang2019quantum,yang2019engineering} and second-order transitions~\cite{zanardi2006ground,zanardi2007mixed,gu2008fidelity,gammelmark2011phase,skotiniotis2015quantum,chu2021dynamic,montenegro2021global,di2023critical}, localization phenomena~\cite{he2023stark,sahoo2024localization,yousefjani2025nonlinearity,sahoo2026enhanced}, and topological transitions~\cite{sarkar2022free,mukhopadhyay2024modular,he2025quantum}, and have been investigated experimentally on a broad range of platforms~\cite{yu2020experimental,liu2021experimental,ding2022enhanced,yu2024experimental,Ilias2024,yu2025experimental,beaulieu2025criticality,lu2026critical}.
For these schemes on quantum platforms, the quantum Fisher information (QFI) with respect to the sensing parameter $\theta$, denoted by $F_\theta$,
quantifies the ultimate achievable precision via the quantum Cram\'er--Rao bound~\cite{braunstein1994statistical,cramer1999mathematical},
and can scale superlinearly with the system-size resource ${N}$, enabling sensitivity beyond the SQL. For continuous critical points, one often finds an algebraic enhancement $F_\theta \sim {N}^{\beta}$, with $\beta>1$ indicating beyond-SQL scaling~\cite{rams2018limits}.

There are however bottlenecks to bring critical quantum metrology closer to practice -- the criticality that enhances system's sensitivity also critically slows down the sensing dynamics and drastically shrinks the workable size of the sensing parameter window. Indeed,  near a continuous quantum phase transition, the energy gap (denoted $\Delta E$) closes in the thermodynamic limit and obeys the universal finite-size scaling $\Delta E\sim {N}^{-z/d}$, where $z$ is the dynamical critical exponent and $d$ is the spatial dimension~\cite{sachdev1999quantum,dziarmaga2010dynamics,polkovnikov2011colloquium}. As a consequence, the gap-induced timescale grows rapidly with ${N}$, leading to critical slowing down and thus diminishing the metrological gain in realistic time-constrained settings~\cite{rams2018limits}. Furthermore, the criticality-based enhancement is effectively restricted to a vanishingly small neighborhood of the critical point, with width $\delta\theta \sim {N}^{-1/(d\nu)}$, where $\nu>0$ is the correlation-length exponent~\cite{sachdev1999quantum,dziarmaga2010dynamics,polkovnikov2011colloquium}. The required tuning precision hence tightens with the increasing system size ${N}$. 

To boost the relevance of critical sensors, it is necessary to invent mechanisms that can retain some critical response features but at the same time mitigate the above-mentioned hurdles~\cite{mukhopadhyay2024modular,xu2025toward,mihailescu2026anti}. In this Letter, we propose such a solution based on  Fock-space lattices (FSLs). We illustrate the physical mechanism using a two-cavity Jaynes--Cummings-type model within a fixed total-excitation sector $N$. In its one-dimensional FSL representation, the ratio of the two qubit--cavity couplings encodes the sensing parameter in the spatial profile of a topological zero-energy mode. Bosonic ladder-operator matrix elements render the effective hoppings cell dependent, so that, as the cell index varies, the effective hopping ratios trace out a curve in an associated topological phase diagram. Cell-dependent criticality (a terminology we introduce here) emerges when part of this curve approaches or crosses a phase boundary, producing a strong sensing response over a broad parameter range. Because only part of the effective lattice is required to approach criticality, our sensing protocol avoids global critical tuning and already suggests a reduced gap cost relative to the homogeneous local-encoding benchmark. Moreover, the multiphoton process reshapes the curve toward a multi-critical point, continuously tuning the QFI scaling from $F_\theta\sim N$ to $F_\theta\sim N^2$ together with the associated sensitivity--gap trade-off. It is also found that the QFI can be saturated by a local photon-number measurement on a single cavity. These results show that intrinsic hopping inhomogeneity in FSLs can alleviate key bottlenecks of critical metrology through cell-dependent criticality.

\begin{figure}[!t]
\centering
\includegraphics[width=\columnwidth]{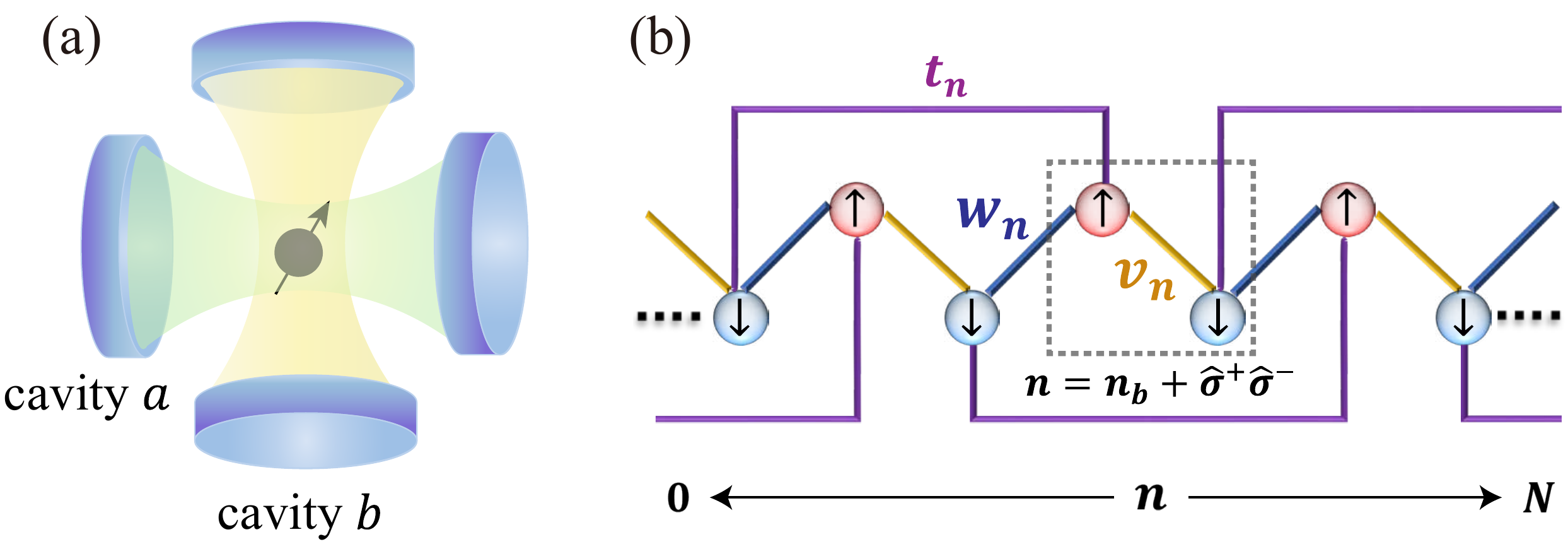}
\caption{\label{fig1}
\textbf{Fock-space lattice representation of the Jaynes--Cummings-type model.} (a) Schematic of the two-cavity Jaynes--Cummings-type setup described by Eq.~\eqref{eq:H}.
(b) Corresponding one-dimensional FSL in the basis $|s,n_a,n_b\rangle$, with $s=\uparrow,\downarrow$ labeling the two sublattices. The hoppings $v_n$, $w_n$, and $t_n$, defined in Eq.~\eqref{eq:Str}, preserve chiral symmetry. The states $\lvert\uparrow,N-n,n-1\rangle$ and $\lvert\downarrow,N-n,n\rangle$ form the $n$-th cell, indicated by the dashed box.}
\end{figure}

{\it Fock-space lattice}.---Bosonic cavity systems naturally accommodate the Fock space as a synthetic dimension~\cite{cai2021topological,pan2021point,deng2022observing,yuan2025quantum,zhang2025synthetic,yao2026non}.  Consider a two-level system resonantly coupled to two bosonic modes $\hat a$ and $\hat b$ [Fig.~\ref{fig1}(a)], described by
\begin{equation}
\hat H=
g\left(\sin\theta\hat a^{\dagger}\hat{\sigma}^-+\cos\theta\hat b^{\dagger}\hat{\sigma}^-\right)
+\frac{\gamma}{N}\hat a^{\dagger}\hat a^{\dagger}\hat b\hat{\sigma}^-
+\mathrm{H.c.},
\label{eq:H}
\end{equation}
where $\hat{\sigma}^-$ is the qubit lowering operator. The total excitation number $\hat{\mathcal N}=\hat a^\dagger \hat a+\hat b^\dagger \hat b+\hat\sigma^+\hat\sigma^-$ is conserved, and we work in a fixed-$N$ sector.  Here the factor $1/N$ accompanying $\gamma$ is a normalization convention that keeps $\gamma$ as a meaningful nonlinear control parameter in the semiclassical limit and enables finite-size scaling analysis. The sensing parameter $\theta$ is encoded in the ratio of the two qubit--cavity couplings, whereas $\gamma$ characterizes the multiphoton process. We set the linear-coupling scale as the energy unit, $g\equiv1$.

\begin{figure}[!b]
\centering
\includegraphics[width=\columnwidth]{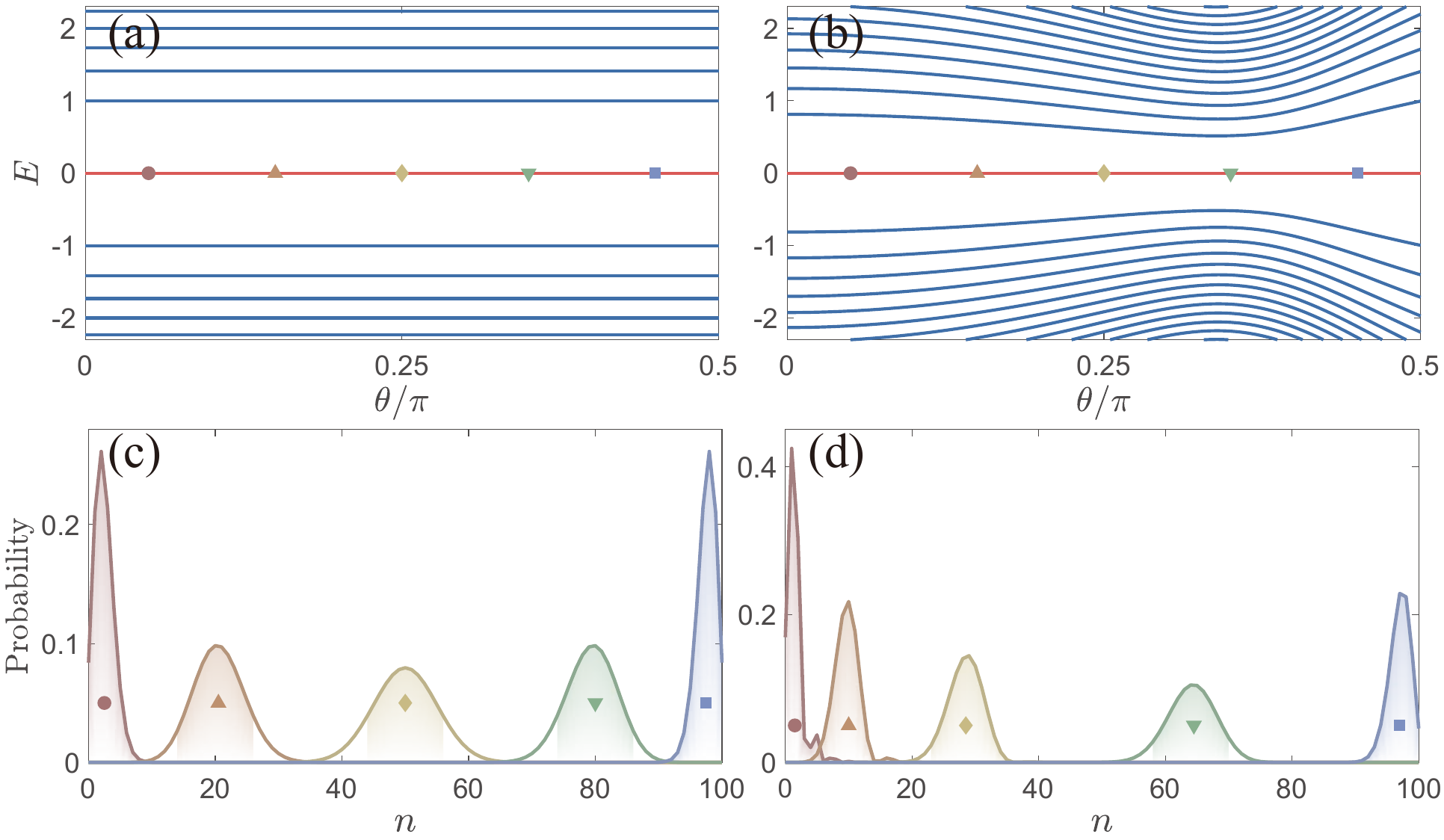}
\caption{\label{fig2}
\textbf{Energy spectra and zero-energy mode profiles of the Fock-space lattice sensor.}
(a),(b) Energy spectra versus $\theta$ for $N=100$ without ($\gamma=0$) and with the nonlinear process ($\gamma=0.6$), respectively. (c),(d) Corresponding distributions of the zero-energy mode $|\psi_0\rangle$ over the FSL cells for representative values of $\theta$, indicated by the colored markers in (a),(b), respectively.
}
\end{figure}

In the fixed-$N$ sector, the $2N+1$ states $|s,n_a,n_b\rangle$ form a one-dimensional FSL, as shown in Fig.~\ref{fig1}(b). Here the qubit states $s=\uparrow,\downarrow$ define the two sublattices, and $n_a$ ($n_b$) denotes the photon number of mode $\hat a$ ($\hat b$). Bosonic ladder operators carry occupation-dependent matrix elements, rendering the effective hoppings along the FSL intrinsically site dependent. The two linear couplings generate nearest-neighbor hoppings, and the multiphoton process adds a longer-range hopping channel:
\begin{equation}
\lvert\downarrow,N-n+j,n-j\rangle \leftrightarrow \lvert\uparrow,N-n,n-1\rangle,\quad j=0,1,2.
\label{eq:FSL}
\end{equation}
The corresponding hopping amplitudes, as specifically indicated on 
Fig.~1(b) are given by
\begin{equation}
\begin{aligned}
v_n &= g\cos\theta\sqrt{n},\qquad
w_n = g\sin\theta\sqrt{N-n+1},\\
t_n &= \frac{\gamma}{N}\sqrt{(n-1)(N-n+1)(N-n+2)}.
\end{aligned}
\label{eq:Str}
\end{equation}
Because every hopping flips the qubit state, the FSL is bipartite and chiral symmetric. Its odd number of sites guarantees an exact zero-energy mode $|\psi_0\rangle$ both at $\gamma=0$ and for finite $\gamma$, as seen from the spectra in Figs.~\ref{fig2}(a,b). Figures~\ref{fig2}(c,d) show the corresponding distributions of $|\psi_0\rangle$ over the FSL cells for representative values of $\theta$; its explicit form is given in the Supplemental Material~\cite{SM}. The strong $\theta$ dependence of the zero-mode distribution motivates the sensing protocol studied below. 

\begin{figure}[!t]
\centering
\includegraphics[width=\columnwidth]{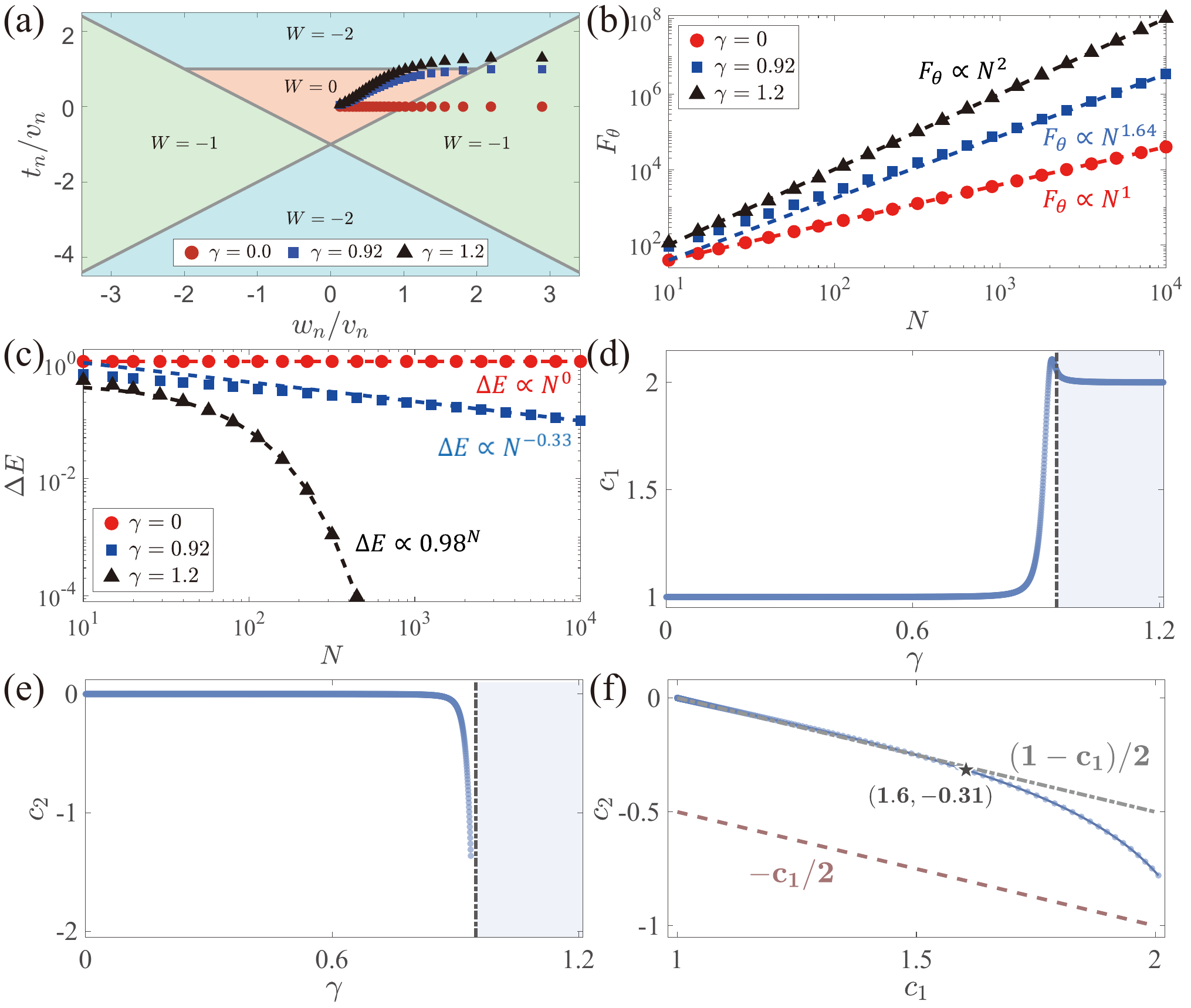}
\caption{\label{fig3}
\textbf{Cell-dependent criticality and the resulting favorable sensitivity-gap trade-off.}
(a) Phase diagram of the associated extended SSH model in the $(w_n/v_n,t_n/v_n)$ plane, with winding numbers $W=0,-1,-2$. The colored curves show representative cell-dependent curves for different $\gamma$ at $\theta=0.2\pi$.
(b) Finite-size scaling of the QFI $F_\theta$, showing progressively enhanced scaling as $\gamma$ increases.
(c) Corresponding excitation gap $\Delta E$ versus $N$, showing a constant gap at $\gamma=0$, algebraic suppression at intermediate $\gamma$, and exponential suppression at larger $\gamma$.
(d),(e) Effective scaling exponents $c_1$ and $c_2$, defined by $F_\theta\sim N^{c_1}$ and $\Delta E\sim N^{c_2}$, versus $\gamma$ along the same cut. As $\gamma$ increases, $c_1$ rises from $1$ and $c_2$ decreases from $0$, until the large-$\gamma$ regime (shaded region), where the gap becomes exponentially small and $c_2$ is no longer well defined.
(f) Parametric relation between $c_1$ and $c_2$ in the practically relevant regime $1<c_1\le 2$. The FSL data lie above the homogeneous local-encoding bound $c_2=-c_1/2$ (brown dashed line), indicating a more favorable trade-off than the benchmark from Eq.~\eqref{eq:homo_bound}. The marked point, $(c_1,c_2)\approx(1.64,-0.33)$, illustrates a representative compromise between sensitivity enhancement and the gap cost.
}

\end{figure}

{\it Mechanism and features of the sensor}.---To elaborate on the sensing mechanism, we group the two states $\lvert\uparrow,N-n,n-1\rangle$ and $\lvert\downarrow,N-n,n\rangle$ into the $n$-th cell, labeled by $n=n_b+\hat{\sigma}^+\hat{\sigma}^-$. The corresponding intracell, nearest-neighbor intercell, and second-neighbor intercell hoppings are the above-defined amplitudes $v_n$, $w_n$, and $t_n$, respectively. To digest the physics associated with such a lattice without translational invariance, we connect the FSL with an ensemble of topological lattices.  That is,  for each fixed $n$, these hopping amplitudes (especially their ratios) {\it locally} define an associated homogeneous extended Su--Schrieffer--Heeger (SSH) model~\cite{su1979solitons}, described by the Bloch Hamiltonian
\begin{equation}
H(k)=\left(v_n+w_n e^{-ik}+t_n e^{-2ik}\right)\hat\sigma^+ + \mathrm{H.c.},
\label{eq:Hopro}
\end{equation}
and winding number~\cite{ryu2010topological,chiu2016classification}
\begin{equation}
W=\frac{1}{2\pi i}\int_{-\pi}^{\pi} dk\partial_k \log\left(v_n+w_n e^{-ik}+t_n e^{-2ik}\right).
\label{eq:Wn}
\end{equation}
The resulting phase diagram contains regions with $W=0,-1,-2$, as shown in Fig.~\ref{fig3}(a). As the cell index $n$ varies, the cell-dependent hopping ratios trace out a curve in this phase diagram, providing a compact geometric representation of the inhomogeneous FSL. A representative example at $\theta=0.2\pi$ is shown in Fig.~\ref{fig3}(a), where each of the plotted curve traced out by a varying $n$ depicts the corresponding parameter ratios and the associated topological invariants for each and every member of the lattice ensemble.  When this curve crosses a phase boundary, the system exhibits cell-dependent criticality: the zero-energy mode becomes highly sensitive to $\theta$ through strong redistribution over the FSL cells, thereby generating a strong sensing response. In the large-$N$ limit, one end of the curve is effectively pinned near the origin in the $W=0$ region, whereas the remaining portion is reshaped by $\gamma$. Tuning $\gamma$ can therefore drive the curve across different phase boundaries, even towards a multi-critical point, thus capable of producing distinct sensing characteristics~\cite{cheng2025super,he2025quantum} if our physical insights into the inhomogeneous FSL are correct. 

To quantify the sensing performance, we focus on two quantities: the QFI $F_\theta$ of the zero-energy mode $|\psi_0\rangle$, which sets the ultimate sensitivity~\cite{paris2009quantum}, and the excitation gap $\Delta E$, which sets the spectral cost,
\begin{equation}
\begin{aligned}
F_\theta &= 4\left[\langle\partial_{\theta}\psi_0|\partial_{\theta}\psi_0\rangle
-|\langle\psi_0|\partial_{\theta}\psi_0\rangle|^2\right],\
\Delta E &= \min_{\mu\neq 0}|E_\mu| ,
\end{aligned}
\label{eq:QFI_gap}
\end{equation}
where ${E_\mu}$ are the eigenenergies of $\hat H$ in the fixed-$N$ sector, and $E_0=0$ corresponds to the zero-energy mode.

We first consider the linear case $\gamma=0$. In this limit, the cell-dependent effective hoppings trace out curves on the $t_n=0$ cut of the phase diagram; a representative example at $\theta=0.2\pi$ is shown in Fig.~\ref{fig3}(a), where the corresponding curve crosses the boundary between $W=0$ and $W=-1$ boundary. This boundary-crossing behavior is seen to yield a sensing response of $F_{\theta}=4N\sim N$ and a constant gap $\Delta E=1\sim N^0$, as shown in Figs.~\ref{fig3}(b,c). This linear-limit behavior holds for all $\theta$: the spectrum in Fig.~\ref{fig2}(a) shows that the gap remains fixed at $\Delta E=1$, with the analytic derivation of $F_\theta=4N$ is detailed elsewhere~\cite{SM}. Increasing the nonlinear coupling parameter $\gamma$ progressively pushes the curve in Fig.~\ref{fig3}(a) toward the multi-critical point, which indeed enhances the sensitivity of the zero-energy mode to $\theta$. For example, near $\gamma\simeq 0.92$, the finite-size scaling in Figs.~\ref{fig3}(b,c) gives $F_\theta\sim N^{1.64}$ together with $\Delta E\sim N^{-0.33}$. For larger $\gamma$, the curve is further deformed and can cross the phase boundary between $W=0$ and $W=-2$. In this regime, the QFI is found to reach the Heisenberg scaling, $F_\theta\sim N^2$, with the excitation gap becoming exponentially small owing to the emergence of additional topological modes. The nonlinear interaction parameter $\gamma$ thus organizes the sensing performance from the robust linear-limit baseline to stronger but more costly protocols.

Taken together, the results above reveal a previously unknown mechanism based on cell-dependent criticality: the intrinsic lattice inhomogeneity with a fixed parameter set can effectively yield an ensemble of homogeneous lattices as illustrated by one curve in the above-presented topological phase diagram. Importantly, only part of the FSL approaches or crosses a phase boundary, thus pointing to a favorable sensitivity-gap trade-off and broad sensing coverage, which we analyze more systematically in the following section.

{\it Sensitivity--gap trade-off}.---We next quantify the sensing performance by investigating the finite-size scaling exponents of the QFI together with that of the excitation gap, namely, 
\begin{equation}
F_\theta \sim N^{c_1},\qquad \Delta E \sim N^{c_2}.
\label{eq:tradeoff_exp}
\end{equation}
Results of $c_1$ and $c_2$ as a function of the nonlinear strength $\gamma$ in the representative case of $\theta=0.2\pi$ are shown in Figs.~\ref{fig3}(d,e). Here, a larger $c_1$ indicates stronger sensitivity enhancement, whereas a more negative $c_2$ indicates a higher gap cost. For small $\gamma$, where the curve crosses only the boundary between $W=0$ and $W=-1$, we obtain $c_1=1$ and $c_2=0$, corresponding to the standard scaling with an $N$-independent gap. For sufficiently large $\gamma$ (shaded region), the curve also crosses the boundary between $W=0$ and $W=-2$. In this regime, the QFI reaches the Heisenberg scaling $c_1=2$, whereas the gap becomes exponentially small, so that a power-law exponent $c_2$ is no longer meaningful. This exponentially small gap comes from additional topological interface modes: in a finite FSL, their hybridization is exponentially weak, yielding an exponentially small splitting. The exact zero mode remains protected, but its adiabatic preparation becomes increasingly susceptible to leakage into these nearby modes.

Between the above-discussed two limiting regimes lies a highly useful working regime, where increasing $\gamma$ raises $c_1$ above the standard-scaling value $1$ but only gradually lowering $c_2$. As shown in Fig.~\ref{fig3}(d), $c_1$ increases from $1$ toward $2$ before the gap enters the exponentially small regime. For $1<c_1\lesssim 2$, $c_2$ varies only mildly, without seeing the gap cost deteriorating rapidly.  We therefore focus on the working regime of $1<c_1\lesssim 2$, whose sensitivity--gap trade-off is summarized in Fig.~\ref{fig3}(f).
For illustration, Fig.~\ref{fig3}(f) marks a representative operating point, $(c_1,c_2)\approx(1.64,-0.33)$, which provides a concrete compromise between sensitivity enhancement and gap cost. The plotted behavior differs qualitatively from conventional critical sensors, where the scaling of both the QFI and and of the gap is typically tied to the universality class of the underlying transition~\cite{montenegro2025quantum}. Indeed, here the nonlinear control parameter $\gamma$ can reshape the ensemble behavior on the topological phase diagram and continuously tune the sensitivity--gap trade-off within a single sensing model.  From above it is now also seen that the intrinsic hopping inhomogeneity  renders it more favorable than in typical homogeneous counterparts with constant hopping amplitudes. The underlying physical reason is rather general: the inhomogeneous FSL supports cell-dependent criticality: only a subset of cells needs to move close to a phase boundary and contribute strongly to the sensing response, whereas the remaining cells stay away from criticality,thus suggesting a reduced gap cost for a given QFI scaling.  By contrast, in homogeneous settings, comparable enhancement typically requires the entire lattice to be tuned close to criticality. 

To digest the physics in more depth we can use a specific case of $\gamma=0$. In this case the FSL already exhibits cell-dependent criticality, the mapped ensemble of the homogeneous lattices crosses of the phase boundary between $W=0$ and $W=-1$ but does not incur any critical slowing down because of the constant gap.  A more general understanding for this advantage can be obtained from a nondegenerate perturbation theory. For a more general sensing protocol, where the state $|E_0\rangle$ responds to a parameter-dependent driving Hamiltonian $\hat H_r$, the QFI admits the spectral representation~\cite{you2007fidelity}
\begin{equation}
F_\theta
=
4\sum_{\mu\neq 0}
\frac{|\langle E_\mu|\partial_\theta \hat H_r|E_0\rangle|^2}
{|E_\mu-E_0|^2}.
\label{eq:Per}
\end{equation}
Using $|E_\mu-E_0|\ge \Delta E$ for all $\mu\neq 0$, together with completeness, one obtains
\begin{equation}
F_\theta
\le
4\,\Delta E^{-2}
\sum_{\mu\neq 0}
|\langle E_\mu|\partial_\theta \hat H_r|E_0\rangle|^2
\le
4\,\|\partial_\theta \hat H_r\|^2\,\Delta E^{-2}.
\label{eq:bound}
\end{equation}
For a homogeneous reference model in which the parameter is encoded through an $N$-independent local driving term, the operator norm $\|\partial_\theta \hat H_r\|$ remains $N$-independent. Under this assumption, Eq.~(\ref{eq:bound}) reduces, at the level of scaling, to
\begin{equation}
F_\theta \le C\,\Delta E^{-2},
\label{eq:homo_bound}
\end{equation}
where $C$ is independent of $N$. Equation~\eqref{eq:homo_bound} will be used only as a benchmark for this class of homogeneous local encodings, where achieving a QFI scaling $F_\theta\sim N^{c_1}$ always requires the gap to close at least as $\Delta E \lesssim N^{-c_1/2}$.
This scaling estimate is shown as the brown dashed line in Fig.~\ref{fig3}(f) and used as a benchmark. Relative to it, the FSL data satisfy $c_2>-c_1/2$, indicating that the same metrological scaling can be achieved with a smaller gap cost than in this homogeneous local-encoding reference. 
This advantage originates from the intrinsic inhomogeneity of the FSL, which generates cell-dependent criticality and leads to a more favorable sensitivity-gap trade-off than the homogeneous local-encoding benchmark.

\begin{figure}[!t]
\centering
\includegraphics[width=\columnwidth]{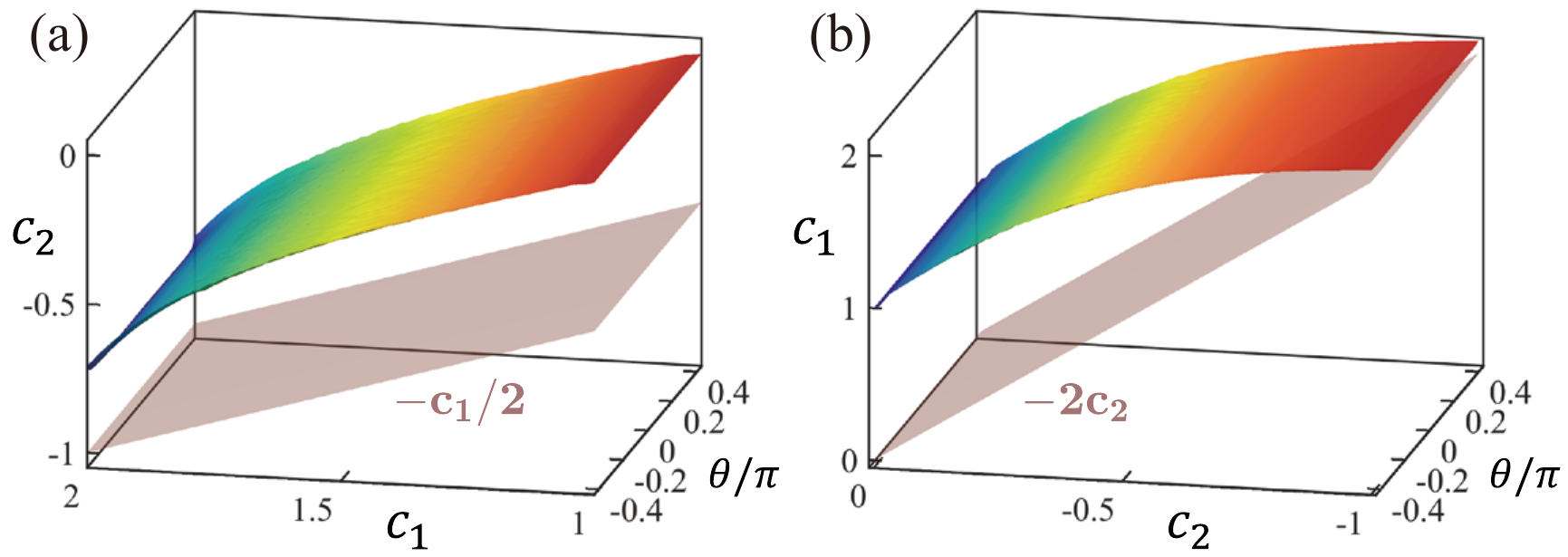}
\caption{\label{fig4}
\textbf{Broad sensing coverage enabled by cell-dependent criticality.}
(a) Gap-cost exponent $c_2$ required to achieve a target QFI scaling exponent $c_1\in(1,2]$, shown versus $\theta/\pi$ for $\gamma>0$.
(b) Achievable QFI scaling exponent $c_1$ under a prescribed gap cost $c_2\in[-1,0]$, also shown versus $\theta/\pi$ for $\gamma>0$.
Together, the two panels show that, even within the practically relevant trade-off regime identified in Fig.~\ref{fig3}(f), the sensing window remains broad in $\theta$.
For comparison, the homogeneous local-encoding benchmark from Eq.~\eqref{eq:homo_bound} is extended over the same $\theta$ interval and shown as the brown shaded surfaces. These surfaces do not represent the actual achievable coverage of a specific homogeneous sensor, since its critical region typically shrinks to a narrow neighborhood of an isolated critical point.
}
\end{figure}

{\it Sensing coverage}.---A further practical advantage of the present FSL sensor is its broad sensing coverage. In conventional critical metrology, the metrological response is typically confined to the vicinity of an isolated critical point~\cite{sachdev1999quantum,dziarmaga2010dynamics,polkovnikov2011colloquium}, resulting in a drastically narrowing operation window as the system size scales up.  Here, by contrast, useful metrological enhancement arises from cell-dependent criticality, which only requires the associated ensemble (the curve plotted on the phase diagram) to cross or approach a phase boundary, a condition much easier to satisfy than tuning the whole homegeous system close to a critical point. As a result, a useful sensing response persists over a finite interval of $\theta$, rather than only at an isolated fine-tuned point.
This feature is already evident in the linear limit $\gamma=0$, where standard scaling persists across the full sensing range. For sufficiently large $|\gamma|$, the curve intersects the boundary between $W=0$ and $W=-2$ for most values of $\theta$ (except at the degenerate points $\theta=\pm \pi/2$, where $v_n=0$), so that the QFI reaches the Heisenberg scaling $F_\theta\sim N^2$ over a broad interval, albeit with an exponentially small gap. More importantly, the practically relevant intermediate regime discussed above can persist over a broad sensing range while retaining a favorable and continuously tunable sensitivity--gap trade-off.

For $\gamma>0$, this practically relevant regime spans approximately $\theta\in[-0.44\pi,\,0.44\pi]$. Figures~\ref{fig4}(a,b) summarize it from two complementary perspectives: Fig.~\ref{fig4}(a) shows, within this interval, the gap-cost exponent $c_2$ required to achieve a target $c_1\in(1,2]$, whereas Fig.~\ref{fig4}(b) shows the achievable $c_1$ for a prescribed $c_2\in[-1,0]$. The brown shaded surfaces show the homogeneous local-encoding benchmark from Eq.~\eqref{eq:homo_bound}, evaluated over the same $\theta$ interval for reference. They are included only as a comparison baseline, not as the actual operating window of a homogeneous sensor. Compared with this benchmark, the FSL results remain systematically superior across this broad $\theta$ interval.
Moreover, the Hamiltonian in Eq.~\eqref{eq:H} satisfies the unitary equivalence
\begin{equation}
U\hat H(\theta,\gamma)U^\dagger=\hat H(\theta+\pi,-\gamma),
\end{equation}
where $U=e^{i\pi \hat\sigma^+\hat\sigma^-}$ flips the sign of $\hat\sigma^-$. Therefore, the $\gamma>0$ window $\theta\in[-0.44\pi,\,0.44\pi]$ maps, for $\gamma<0$, to $\theta\in[-\pi,\,-0.56\pi]\cup[0.56\pi,\,\pi]$. In the remaining regions, except at $\theta=\pm\pi/2$, increasing $|\gamma|$ can also raise $c_1$ from $1$ toward $2$, but the gap often deteriorates rapidly and may even become exponentially small before a super-SQL regime is reached. The Supplemental Material~\cite{SM} further analyzes these complementary regions and estimates the practical sensing window, consistent with the picture of cell-dependent criticality developed above. This picture explains why the sensing coverage is broad and how the practical parameter window is determined.

{\it State preparation and read-out}.---A simple protocol starts from $\lvert\downarrow,N,0\rangle=\lvert\downarrow\rangle\otimes\lvert N\rangle_a\otimes\lvert0\rangle_b$, which is the exact zero-energy mode at $\theta=\gamma=0$. Large Fock-state preparation with $O(10^2)$ photons has already been demonstrated in circuit-QED platforms~\cite{deng2024quantum,xu2026principles}, and scalable proposals further extend this capability to even larger photon numbers~\cite{li2026scalable}. One may first keep $\gamma=0$ and tune $\theta$ to the target value, where the gap to finite-energy states remains fixed at $\Delta E=1$, and then ramp up $|\gamma|$ into the desired operating regime. Along this ramp, the QFI scaling is continuously enhanced from $F_\theta\sim N$ toward $F_\theta\sim N^2$ within the practically relevant trade-off regime, which correspondingly relaxes the precision required for tuning $\gamma$. A possible circuit-QED implementation based on longitudinal-coupling-induced multiphoton processes~\cite{zhao2015generating,zhao2016engineering} is outlined in the Supplemental Material~\cite{SM}. Moreover, because the zero-energy mode has support only on the $\downarrow$ sublattice whereas finite-energy eigenstates occupy both sublattices, any detected $\uparrow$-sector population directly signals diabatic leakage during preparation. After preparation, the QFI can already be saturated, in the effective sensing frame, by a local photon-number measurement on a single cavity~\cite{schuster2007resolving,lu2023resolving}, since the real hoppings in Eq.~\eqref{eq:Str} allow the zero-energy-mode amplitudes to be chosen real in the Fock basis~\cite{sarkar2022free}; see the Supplemental Material~\cite{SM}. In the laboratory frame, this corresponds to a dressed local observable involving the same cavity mode and the qubit, which reduces to standard photon-number-resolved readout up to perturbatively small corrections for weak dressing, or can be mapped back explicitly by undoing the dressing transformation before measurement.

{\it Conclusion}.---To mitigate current bottlenecks in applying critical quantum metrology, we advocate in this work cell-dependent criticality through the use of FSLs.  The intrinsic site-dependent hoppings in a FSL can efectively map one inhomogeneous lattice to an ensemble of homogeneous lattices and hence trace out one curve on the associated topological phase diagram.  Critical sensing emerges when this curve crosses a phase boundary, and this only requires part of the lattice to approach or cross a phase boundary. Such cell-dependent criticality yields a continuously tunable sensitivity-gap trade-off that is more favorable than in a homogeneous local-encoding scenario  while retaining broad sensing coverage.  We conclude that FSLs not only serve as versatile platforms for synthetic quantum matter, but also work as scalable architectures for practical quantum metrology.

\clearpage

\onecolumngrid

\begin{center}
\textbf{\large Supplementary Material for ``Cell-Dependent Criticality for Quantum Metrology''}
\end{center}

\tableofcontents
\setcounter{secnumdepth}{2}

		%%%%%%%%%% Prefix a "S" to all equations, figures, tables and reset the counter %%%%%%%%%%
	\setcounter{equation}{0} \setcounter{figure}{0} \setcounter{table}{0} %
	\renewcommand{\theequation}{S\arabic{equation}} \renewcommand{\thefigure}{S%
		\arabic{figure}} \renewcommand{\bibnumfmt}[1]{[S#1]} 
        \newcommand\underrel[2]{\mathrel{\mathop{#2}\limits_{#1}}}
	%\renewcommand{\citenumfont}[1]{S#1}
	%%%%%%%%%% Prefix a "S" to all equations, figures, tables and reset the counter %%%%%%%%%%

\section{Derivation of the topological zero-energy mode}\label{SM1}

In the fixed-$N$ excitation sector, the Hamiltonian in Eq.~(1) of the main text is mapped to a Fock-space lattice (FSL) with $2N+1$ basis states,
\[
\lvert\downarrow,N-n,n\rangle,\quad n=0,1,\dots,N,
\]
and
\[
\lvert\uparrow,N-n,n-1\rangle,\quad n=1,2,\dots,N.
\]
Since every term in the Hamiltonian flips the qubit state, the resulting FSL is bipartite and obeys chiral symmetry. Moreover, the two sublattices contain $N+1$ and $N$ sites, respectively, so the sublattice imbalance guarantees at least one exact topological zero-energy mode. In the present case, this zero-energy mode resides entirely on the $\downarrow$ sublattice, and can therefore be written as
\begin{equation}
|\psi_0\rangle=\sum_{n=0}^{N} u_n\,\lvert\downarrow,N-n,n\rangle .
\label{eq:SM_zero_mode_general}
\end{equation}
Its explicit form follows from the destructive-interference conditions on all $\uparrow$-sublattice sites, namely
\begin{equation}
\langle \uparrow,N-n,n-1|\hat H|\psi_0\rangle=0,
\label{eq:rel}
\end{equation}
which yield the recursion relations for the amplitudes $\{u_n\}$.

For the linear case $\gamma=0$, Eq.~\eqref{eq:rel} reduces to
\begin{equation}
\cos\theta\,\sqrt{n}\,u_n+\sin\theta\,\sqrt{N-n+1}\,u_{n-1}=0,
\qquad n=1,2,\dots,N.
\label{eq:SM_recursion_linear}
\end{equation}
This first-order recursion can be solved analytically, yielding the normalized zero-energy mode
\begin{equation}
|\psi_0\rangle
=
\sum_{n=0}^{N}
\sqrt{\frac{N!}{n!(N-n)!}}\,
(-\sin\theta)^n(\cos\theta)^{N-n}
\,\lvert\downarrow,N-n,n\rangle .
\label{eq:SM_zero_mode_linear}
\end{equation}
This is a two-mode binomial state, which may also be viewed as an SU(2) coherent state in the fixed-$N$ sector. Accordingly, the probability distribution
\begin{equation}
P_n=|\langle \downarrow,N-n,n|\psi_0\rangle|^2
=
\binom{N}{n}
(\sin^2\theta)^n(\cos^2\theta)^{N-n}
\label{eq:SM_binomial_distribution}
\end{equation}
is binomial, with mean and variance
\begin{equation}
\langle n\rangle = N\sin^2\theta,
\qquad
\mathrm{Var}(n)=N\sin^2\theta\cos^2\theta .
\label{eq:SM_mean_variance}
\end{equation}
Therefore, varying $\theta$ continuously shifts the center of the zero-energy-mode distribution along the FSL and also changes its width, as shown in Fig.~2(c) of the main text. In particular, the state is localized at $n=0$ for $\theta=0$, at $n=N$ for $\theta=\pi/2$, and is most broadly distributed around $\theta=\pi/4$. This strong and controllable redistribution of the zero-energy mode underlies its sensing capability with respect to $\theta$.

For $\gamma\neq 0$, the zero-energy mode no longer admits an equally compact closed-form expression. Instead, it is determined by the recursion relations associated with the cell-dependent hoppings, reproduced here from Eq.~(3) of the main text for convenience:
\begin{equation}
\begin{aligned}
v_n &= g\cos\theta\sqrt{n},\qquad
w_n = g\sin\theta\sqrt{N-n+1},\\
t_n &= \frac{\gamma}{N}\sqrt{(n-1)(N-n+1)(N-n+2)}.
\end{aligned}
\label{eq:SM_hoppings}
\end{equation}
Substituting Eq.~\eqref{eq:SM_zero_mode_general} into the destructive-interference condition Eq.~\eqref{eq:rel}, we obtain
\begin{equation}
v_n u_n + w_n u_{n-1} + t_n u_{n-2}=0,
\qquad n=1,2,\dots,N,
\label{eq:SM_recursion_general}
\end{equation}
with $u_{-1}=0$, while the overall scale is fixed by normalization. This yields $N$ equations for the $N+1$ amplitudes $\{u_n\}_{n=0}^N$, leaving one overall degree of freedom, consistent with the sublattice imbalance that guarantees the existence of the zero-energy mode. Once this zero mode is determined recursively, one sees that the nonlinear process reshapes its distribution along the FSL, as illustrated in Fig.~2(d) of the main text, and thereby modifies its response to $\theta$ and the associated sensing characteristics.

\section{Analytic derivation of \texorpdfstring{$F_\theta=4N$}{Fθ=4N} in the linear limit}\label{SM2}

In the main text, we showed that in the linear case $\gamma=0$, the zero-energy mode exhibits the standard scaling
\[
F_\theta=4N,
\]
independent of $\theta$. This $\theta$-independent linear-limit result provides a broad sensing baseline and sets the stage for the nonlinear enhancement discussed in the main text. In this section, we present two complementary derivations of this result: one based directly on the explicit zero-energy-mode wavefunction derived in Sec.~\ref{SM1}, and the other based on a collective-mode decomposition of the linear Hamiltonian.

Using the linear-case Fock-basis amplitudes $u_n$ derived in Sec.~\ref{SM1},
\begin{equation}
u_n=
\sqrt{\frac{N!}{n!(N-n)!}}\,
(-\sin\theta)^n(\cos\theta)^{N-n},
\label{eq:SM_un_linear_repeat}
\end{equation}
we directly calculate the QFI from
\begin{equation}
F_\theta
=
4\left(
\langle \partial_\theta\psi_0|\partial_\theta\psi_0\rangle
-
|\langle \psi_0|\partial_\theta\psi_0\rangle|^2
\right).
\label{eq:SM_QFI_def}
\end{equation}
Since the amplitudes $u_n$ are real, Eq.~\eqref{eq:SM_zero_mode_linear} gives
\begin{equation}
\langle \psi_0|\partial_\theta\psi_0\rangle
=
\sum_{n=0}^{N} u_n\,\partial_\theta u_n
=
\frac{1}{2}\partial_\theta \sum_{n=0}^{N} u_n^2
=
0,
\label{eq:SM_overlap_zero}
\end{equation}
because the state is normalized.

Next, differentiating Eq.~\eqref{eq:SM_un_linear_repeat} gives
\begin{equation}
\partial_\theta u_n
=
u_n\left(n\cot\theta-(N-n)\tan\theta\right)
=
u_n\,\frac{n-N\sin^2\theta}{\sin\theta\cos\theta}.
\label{eq:SM_dun}
\end{equation}
Substituting Eq.~\eqref{eq:SM_dun} into Eq.~\eqref{eq:SM_QFI_def}, we obtain
\begin{equation}
F_\theta
=
4\sum_{n=0}^{N}(\partial_\theta u_n)^2
=
\frac{4}{\sin^2\theta\cos^2\theta}
\sum_{n=0}^{N}
u_n^2\bigl(n-N\sin^2\theta\bigr)^2.
\label{eq:SM_QFI_direct}
\end{equation}

Now note that $u_n^2=P_n$ is exactly the binomial distribution given in Eq.~\eqref{eq:SM_binomial_distribution}. Using Eq.~\eqref{eq:SM_mean_variance}, we have
\begin{equation}
\sum_{n=0}^{N} u_n^2\,n = \langle n\rangle = N\sin^2\theta.
\label{eq:SM_mean_n_repeat}
\end{equation}
Therefore, the summation in Eq.~\eqref{eq:SM_QFI_direct} is precisely the variance,
\begin{equation}
\sum_{n=0}^{N}
u_n^2\bigl(n-N\sin^2\theta\bigr)^2
=
\sum_{n=0}^{N} u_n^2\bigl(n-\langle n\rangle\bigr)^2
=
\mathrm{Var}(n)
=
N\sin^2\theta\cos^2\theta.
\label{eq:SM_variance_identity}
\end{equation}
It then follows that
\begin{equation}
F_\theta=4N.
\label{eq:SM_QFI_4N}
\end{equation}
This proves that the linear-limit QFI is independent of $\theta$ and scales linearly with $N$.

The same result can also be understood more transparently from a collective-mode picture~\cite{cai2021topological,deng2022observing}. For $\gamma=0$, the Hamiltonian in Eq.~(1) of the main text becomes
\begin{equation}
\hat H_{\rm lin}
=
g\left(\sin\theta\,\hat a^\dagger\hat\sigma^-+\cos\theta\,\hat b^\dagger\hat\sigma^-+\mathrm{H.c.}\right).
\label{eq:SM_Hlin}
\end{equation}
We define the bright and dark bosonic modes
\begin{equation}
\hat c^\dagger=\sin\theta\,\hat a^\dagger+\cos\theta\,\hat b^\dagger,
\qquad
\hat d^\dagger=\cos\theta\,\hat a^\dagger-\sin\theta\,\hat b^\dagger,
\label{eq:SM_cd_modes}
\end{equation}
which satisfy canonical bosonic commutation relations and are mutually orthogonal. In terms of these collective modes, the Hamiltonian reduces to the single-mode Jaynes--Cummings form
\begin{equation}
\hat H_{\rm lin}
=
g\left(\hat c^\dagger\hat\sigma^-+\hat c\,\hat\sigma^+\right),
\label{eq:SM_Hlin_c}
\end{equation}
while the dark mode $\hat d$ is completely decoupled from the qubit.

It is convenient to introduce the collective-mode Fock basis
\begin{equation}
|n_c,n_d\rangle_{cd}
=
\frac{(\hat c^\dagger)^{n_c}(\hat d^\dagger)^{n_d}}{\sqrt{n_c!\,n_d!}}|0,0\rangle_{cd},
\label{eq:SM_cd_basis}
\end{equation}
where $n_c$ and $n_d$ denote the occupation numbers of the bright and dark modes, respectively. In this basis, the exact zero-energy mode in the fixed-$N$ sector is simply
\begin{equation}
|\psi_0\rangle
=
|\downarrow\rangle\otimes \frac{(\hat d^\dagger)^N}{\sqrt{N!}}|0,0\rangle_{cd}
=
|\downarrow\rangle\otimes |0,N\rangle_{cd},
\label{eq:SM_dark_mode_state}
\end{equation}
which is manifestly a dark state of $\hat H_{\rm lin}$. Expanding Eq.~\eqref{eq:SM_dark_mode_state} in the original $(a,b)$ Fock basis immediately reproduces Eq.~\eqref{eq:SM_zero_mode_linear}.

This collective-mode representation also gives a simple derivation of the QFI. Since
\begin{equation}
\partial_\theta \hat d^\dagger
=
-\sin\theta\,\hat a^\dagger-\cos\theta\,\hat b^\dagger
=
-\hat c^\dagger,
\label{eq:SM_ddtheta}
\end{equation}
we have
\begin{equation}
\partial_\theta |\psi_0\rangle
=
-\sqrt{N}\,
|\downarrow\rangle\otimes
\hat c^\dagger
\frac{(\hat d^\dagger)^{N-1}}{\sqrt{(N-1)!}}|0,0\rangle_{cd}
=
-\sqrt{N}\,|\downarrow\rangle\otimes |1,N-1\rangle_{cd}.
\label{eq:SM_dpsi_collective}
\end{equation}
Equation~\eqref{eq:SM_dpsi_collective} is orthogonal to Eq.~\eqref{eq:SM_dark_mode_state}, since the former contains one bright-mode excitation whereas the latter contains none. Hence
\begin{equation}
\langle \psi_0|\partial_\theta\psi_0\rangle=0,
\qquad
\langle \partial_\theta\psi_0|\partial_\theta\psi_0\rangle=N.
\label{eq:SM_overlap_collective}
\end{equation}
Substituting these into the definition of the QFI yields
\begin{equation}
F_\theta
=
4\left(
\langle \partial_\theta\psi_0|\partial_\theta\psi_0\rangle
-
|\langle \psi_0|\partial_\theta\psi_0\rangle|^2
\right)
=
4N,
\label{eq:SM_QFI_collective}
\end{equation}
in agreement with Eq.~\eqref{eq:SM_QFI_4N}. This collective-mode picture shows that the $\theta$ dependence enters only through the rotation between the bright and dark modes, while the metrological response remains exactly linear in $N$ throughout the entire linear regime.

\begin{figure}[!b]
\centering
\includegraphics[width=0.9\columnwidth]{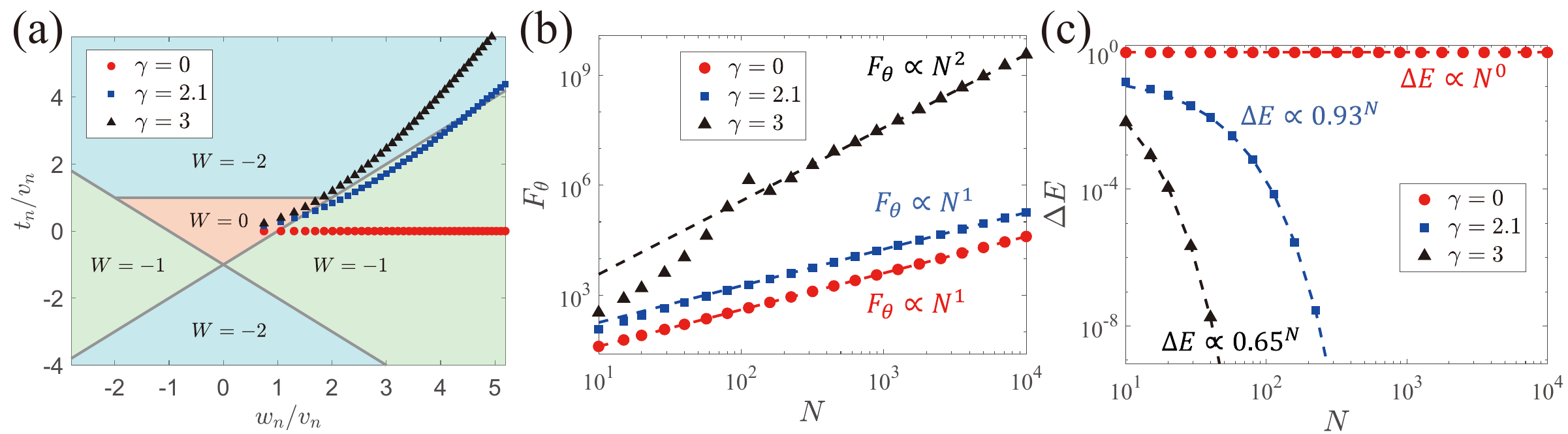}
\caption{\label{figs1}
\textbf{Sensing characteristics of the Fock-space-lattice sensor in an unfavorable parameter region.}
(a) Phase diagram of the associated extended SSH model in the $(w_n/v_n,t_n/v_n)$ plane, with winding numbers $W=0,-1,-2$. The colored curves show representative ones generated by the cell-dependent hoppings at $\theta=0.47\pi$ for different $\gamma$. As $\gamma$ increases, the curve bends upward; for the intermediate case $\gamma=2.1$, it already develops an additional crossing of the $W=-1$--$W=-2$ boundary before reaching the multi-critical point. 
(b) Finite-size scaling of the QFI $F_\theta$. At $\gamma=0$, the QFI exhibits the standard scaling $F_\theta\sim N$. For the intermediate case $\gamma=2.1$, the QFI still remains close to the standard scaling, whereas for the larger value $\gamma=3$, it reaches the Heisenberg scaling $F_\theta\sim N^2$. 
(c) Corresponding excitation gap $\Delta E$ versus $N$. The gap remains constant at $\gamma=0$, but already becomes exponentially small for $\gamma=2.1$ due to the additional topological interface, and is further suppressed for $\gamma=3$.}
\end{figure}

\section{Unfavorable region and geometric estimate of the practical boundary}\label{SM3}

Here we provide further details on the unfavorable parameter regions referred to in the main text, where increasing $|\gamma|$ can in principle drive the QFI exponent $c_1$ from $1$ toward $2$, but severe gap degradation already sets in before the super-SQL regime $c_1>1$ is reached. For definiteness, it is sufficient to focus on the $\gamma>0$ sector and $\theta\in(0,\pi/2)$. Indeed, a $\pi$ phase shift on mode $\hat a$, generated by $U_a=e^{i\pi \hat a^\dagger \hat a}$, gives
\begin{equation}
U_a \hat H(\theta,\gamma) U_a^\dagger=\hat H(-\theta,\gamma),
\label{eq:SM_theta_mirror}
\end{equation}
so the $\theta<0$ side follows by mirror symmetry. The extension to the remaining parameter regions, including the $\gamma<0$ sector, then follows from the symmetry relations discussed in the main text.

As an illustrative example slightly outside the practically relevant window identified in the main text, we consider $\theta=0.47\pi$. As before, the sensing behavior can be understood from the phase-diagram curve traced out by the effective hoppings. In the large-$N$ limit, one end of the curve remains pinned near the origin, while the remaining part bends upward as $\gamma$ increases; see Fig.~\ref{figs1}(a). For $\gamma=0$ or sufficiently small $\gamma$, the curve crosses only the $W=0$--$W=-1$ boundary, yielding the standard scaling of the QFI together with a constant excitation gap. For sufficiently large $\gamma$ [here illustrated by $\gamma=3$ in Fig.~\ref{figs1}(a)], the curve also crosses the $W=0$--$W=-2$ boundary, leading to Heisenberg scaling of the QFI but an exponentially small gap, as shown in Figs.~\ref{figs1}(b,c).

\begin{figure}[!t]
\centering
\includegraphics[width=0.6\columnwidth]{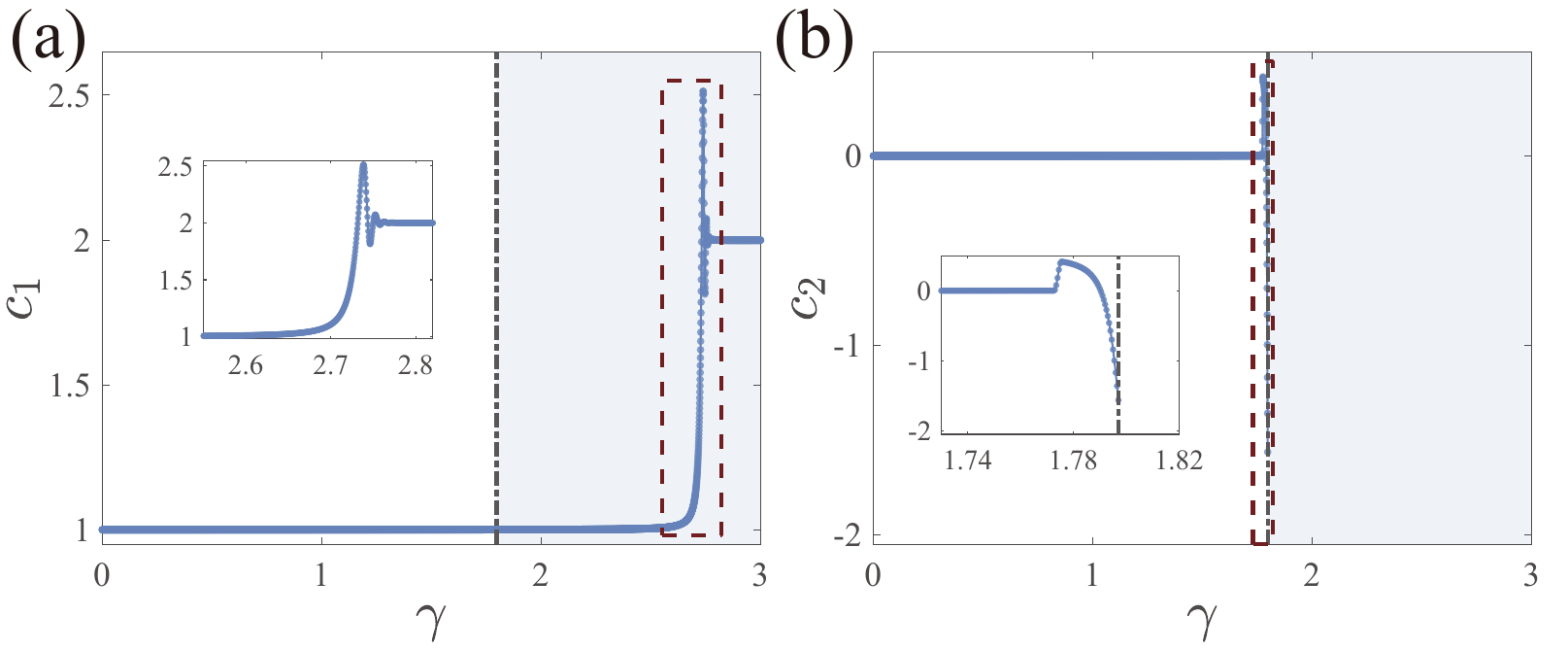}
\caption{\label{figs2}
\textbf{Effective finite-size scaling exponents in an unfavorable parameter region.}
(a),(b) Effective exponents $c_1$ and $c_2$, defined through the finite-size behaviors $F_\theta\sim N^{c_1}$ and $\Delta E\sim N^{c_2}$, versus the nonlinear strength $\gamma$ at $\theta=0.47\pi$. The gray dash-dotted line marks the onset of the exponentially small-gap regime at $\gamma\simeq1.8$, while the shaded region indicates the corresponding large-$\gamma$ side where $c_2$ is no longer well defined. The red dashed boxes mark the ranges enlarged in the insets. As shown in the inset of (a), the QFI exponent remains close to the standard-scaling value $c_1=1$ until $\gamma\simeq2.6$, and only then starts to increase substantially. By contrast, panel (b) and its inset show that severe gap degradation already sets in near $\gamma\simeq1.8$. This identifies an unfavorable intermediate region in which severe gap degradation precedes appreciable metrological enhancement.}
\end{figure}

The key difference from the representative case discussed in the main text is that, for $\theta=0.47\pi$, the upward-bending curve can enter an unfavorable intermediate regime before it reaches the multi-critical point. In this regime, the curve has not yet approached the point closely enough to enhance the QFI beyond the standard scaling, but it already develops an additional crossing of the $W=-1$--$W=-2$ boundary, as illustrated by the $\gamma=2.1$ curve in Fig.~\ref{figs1}(a). This additional crossing indicates the emergence of an extra topological interface and the associated near-zero energy modes, so that the excitation gap already becomes exponentially small even though the QFI still exhibits the standard scaling. To quantify this feature, in Fig.~\ref{figs2} we plot the effective exponents $c_1$ and $c_2$, defined through the finite-size behaviors $F_\theta\sim N^{c_1}$ and $\Delta E\sim N^{c_2}$, as functions of $\gamma$ for the same value $\theta=0.47\pi$. One finds that the gap enters the exponentially small regime already for $\gamma\gtrsim1.8$, whereas the QFI exponent remains close to $c_1=1$ until $\gamma\gtrsim2.6$. This clearly identifies a parameter region in which severe gap degradation sets in before any substantial metrological enhancement is achieved, and hence explains why such a region is not practically relevant.

This example motivates a geometric estimate of the practical boundary. The basic idea is that the favorable window ends when the curve first develops an additional contact with the $W=-1$--$W=-2$ boundary before it can reach the multi-critical point. We now derive this estimate in the large-$N$ limit.

In the large-$N$ limit, it is convenient to introduce the continuum variable
\begin{equation}
s=\frac{n}{N}\in(0,1).
\label{eq:SM_s_def}
\end{equation}
Using the hopping amplitudes in Eq.~(3) of the main text, the curve in the phase diagram can be written as
\begin{equation}
x(s)\equiv \frac{w_n}{v_n}
\approx
\tan\theta\sqrt{\frac{1-s}{s}},
\qquad
y(s)\equiv \frac{t_n}{v_n}
\approx
\frac{\gamma}{\cos\theta}(1-s).
\label{eq:SM_xy_s}
\end{equation}
Eliminating $s$, one obtains the large-$N$ curve
\begin{equation}
y(x)\approx
\frac{\gamma}{\cos\theta}\,
\frac{x^2}{x^2+\tan^2\theta}.
\label{eq:SM_yx}
\end{equation}

The practically relevant region is controlled by the competition between two characteristic events. The first is that the curve reaches the multi-critical point at
\begin{equation}
(x,y)=(2,1).
\label{eq:SM_junction_point}
\end{equation}
Substituting $x=2$ and $y=1$ into Eq.~\eqref{eq:SM_yx}, the corresponding threshold is
\begin{equation}
\gamma_J(\theta)
=
\frac{1+3\cos^2\theta}{4\cos\theta}.
\label{eq:SM_gammaJ}
\end{equation}

The second event is that, before reaching the point, the curve develops an additional contact with the phase boundary
\begin{equation}
y=x-1,
\qquad x>2,
\label{eq:SM_boundary_xminus1}
\end{equation}
which separates the $W=-1$ and $W=-2$ regions. This extra contact signals the appearance of an additional topological interface and hence an exponentially small gap, even though the QFI enhancement may still remain weak. The onset of this unfavorable situation is determined by the tangency conditions
\begin{equation}
y(x_t)=x_t-1,
\qquad
\frac{dy}{dx}\Big|_{x=x_t}=1,
\label{eq:SM_tangent_conditions}
\end{equation}
with $x_t>2$.

Using Eq.~\eqref{eq:SM_yx}, the derivative is
\begin{equation}
\frac{dy}{dx}
=
\frac{\gamma}{\cos\theta}
\frac{2x\tan^2\theta}{(x^2+\tan^2\theta)^2}.
\label{eq:SM_dydx}
\end{equation}
Combining Eq.~\eqref{eq:SM_tangent_conditions} with Eqs.~\eqref{eq:SM_yx} and \eqref{eq:SM_dydx}, one finds
\begin{equation}
\tan^2\theta=\frac{x_t^3}{x_t-2},
\label{eq:SM_tan2_xt}
\end{equation}
and
\begin{equation}
\gamma_t(\theta)
=
2\cos\theta\,\frac{(x_t-1)^2}{x_t-2}.
\label{eq:SM_gammat}
\end{equation}
Here $\gamma_t(\theta)$ is the threshold at which the extra $W=-2$ contact first appears.

The practical boundary is then determined by the condition that the curve reaches the multi-critical point exactly when this additional tangency sets in, namely
\begin{equation}
\gamma_t(\theta_c)=\gamma_J(\theta_c).
\label{eq:SM_critical_condition}
\end{equation}
Substituting Eqs.~\eqref{eq:SM_gammaJ} and \eqref{eq:SM_gammat}, and using Eq.~\eqref{eq:SM_tan2_xt}, this condition reduces to
\begin{equation}
8(x_t-1)^2=x_t^3+4x_t-8,
\label{eq:SM_xt_equation}
\end{equation}
whose relevant solution is
\begin{equation}
x_t=4.
\label{eq:SM_xt4}
\end{equation}
Equation~\eqref{eq:SM_tan2_xt} then gives
\begin{equation}
\tan^2\theta_c=\frac{4^3}{4-2}=32,
\label{eq:SM_tan2_thetac}
\end{equation}
namely
\begin{equation}
\theta_c=\arctan(4\sqrt2)\approx 0.4443\pi.
\label{eq:SM_thetac_final}
\end{equation}

Therefore, for $\theta<\theta_c$, the curve can approach the multi-critical point before an additional $W=-2$ contact induces severe gap degradation, yielding the practically relevant regime discussed in the main text. By contrast, for $\theta>\theta_c$, the curve first develops an extra $W=-1$--$W=-2$ interface, so that the gap can already become exponentially small before substantial QFI enhancement sets in. This provides the geometric explanation for the statement in the main text that, in the remaining parameter regions, severe gap degradation can occur before the super-SQL regime is reached. It also explains why the numerically identified practical window for $\gamma>0$, approximately $\theta\in[-0.44\pi,\,0.44\pi]$, lies slightly inside the geometric estimate \eqref{eq:SM_thetac_final}.

\section{Circuit-QED implementation via longitudinal-coupling-induced multiphoton processes}\label{SM4}

A possible circuit-QED implementation of the Hamiltonian in Eq.~(1) of the main text can be constructed following the longitudinal-coupling scheme developed for superconducting qubits with broken inversion symmetry and its extension to two-cavity systems~\cite{zhao2015generating,zhao2016engineering}. The key idea is that, in addition to the usual transverse interaction, the qubit can also possess longitudinal couplings to the cavity fields. In the presence of a classical drive, these longitudinal couplings generate controllable multiphoton sideband processes, from which the desired linear and nonlinear terms in Eq.~(1) can be isolated.

To keep the notation consistent with the main text, we denote the two cavity modes directly by $\hat a$ and $\hat b$. The starting driven longitudinal-coupling Hamiltonian is
\begin{equation}
\hat H=\hat H_q+\hat H_r+\hat H_g+\hat H_d,
\label{eq:SM_exp_full}
\end{equation}
with
\begin{equation}
\hat H_q=\frac{\omega_x}{2}\hat\sigma_x+\frac{\omega_z}{2}\hat\sigma_z,
\qquad
\hat H_r=\omega_a \hat a^\dagger \hat a+\omega_b \hat b^\dagger \hat b,
\label{eq:SM_exp_qr}
\end{equation}
\begin{equation}
\hat H_g=g_a \hat\sigma_z(\hat a^\dagger+\hat a)+g_b \hat\sigma_z(\hat b^\dagger+\hat b),
\qquad
\hat H_d=\Omega \hat\sigma_z \cos(\tilde\omega t+\phi).
\label{eq:SM_exp_gd}
\end{equation}
Here $g_a$ and $g_b$ are the longitudinal coupling strengths of the qubit to the two cavity modes, while $\tilde\omega$ and $\Omega$ are the frequency and amplitude of the classical drive.

Following Ref.~\cite{zhao2016engineering}, one first performs the displacement transformation
\begin{equation}
\hat D=\exp\!\left[\frac{\eta_a}{2}\hat\sigma_z(\hat a^\dagger-\hat a)+\frac{\eta_b}{2}\hat\sigma_z(\hat b^\dagger-\hat b)\right],
\qquad
\eta_a=\frac{2g_a}{\omega_a},\quad
\eta_b=\frac{2g_b}{\omega_b},
\label{eq:SM_exp_D}
\end{equation}
and then the drive-induced unitary transformation
\begin{equation}
\hat U_d(t)=\exp\!\left[i\frac{x}{2}\hat\sigma_z \sin(\tilde\omega t+\phi)\right],
\qquad
x=\frac{2\Omega}{\tilde\omega}.
\label{eq:SM_exp_Ud}
\end{equation}
After these two steps, the interaction Hamiltonian can be expanded into drive-assisted sideband processes weighted by Bessel functions $J_N(x)$. In the interaction picture, the resonant terms satisfy the standard frequency-matching condition
\begin{equation}
N\tilde\omega+\omega_z+(m_a-n_a)\omega_a+(m_b-n_b)\omega_b=0.
\label{eq:SM_exp_res}
\end{equation}

To realize the model studied in the main text, we choose the resonance conditions
\begin{equation}
\omega_a+\tilde\omega=\omega_z,
\qquad
\omega_b+2\tilde\omega=\omega_z,
\label{eq:SM_exp_match}
\end{equation}
which imply
\begin{equation}
2\omega_a-\omega_b=\omega_z.
\label{eq:SM_exp_match2}
\end{equation}
For $\eta_a,\eta_b\ll 1$, $\phi=0$, and a moderate drive amplitude such that higher-order Bessel terms are negligible, the dominant resonant processes are an $N=-1$ sideband generating a single-photon transition in mode $\hat a$, an $N=-2$ sideband generating a single-photon transition in mode $\hat b$, and an $N=0$ process generating the three-body transition $\hat a^2\hat b^\dagger\hat\sigma_+$. Keeping only the lowest-order contributions in $\eta_a$ and $\eta_b$, one obtains
\begin{equation}
\hat H_{\rm int}
\simeq
\alpha_0\,\hat a\,\hat\sigma_+
+
\beta_0\,\hat b\,\hat\sigma_+
+
\bar\gamma\,\hat a^2\hat b^\dagger\,\hat\sigma_+
+\mathrm{H.c.},
\label{eq:SM_exp_hint}
\end{equation}
with
\begin{equation}
\begin{aligned}
\alpha_0&=-\frac{\omega_x}{2}e^{-(\eta_a^2+\eta_b^2)/2}J_{-1}(x)\eta_a,\\
\beta_0&=-\frac{\omega_x}{2}e^{-(\eta_a^2+\eta_b^2)/2}J_{-2}(x)\eta_b,\\
\bar\gamma&=\frac{\omega_x}{4}e^{-(\eta_a^2+\eta_b^2)/2}J_0(x)\eta_a^2\eta_b.
\end{aligned}
\label{eq:SM_exp_coeff}
\end{equation}

Up to removable phase conventions of the cavity and qubit operators, Eq.~\eqref{eq:SM_exp_hint} is equivalently written as
\begin{equation}
\hat H_{\rm eff}
=
\alpha_0\,\hat a^\dagger \hat\sigma^-
+\beta_0\,\hat b^\dagger \hat\sigma^-
+\bar\gamma\,\hat a^\dagger\hat a^\dagger\hat b\,\hat\sigma^-
+\mathrm{H.c.},
\label{eq:SM_exp_targetbare}
\end{equation}
which already has the same operator structure as Eq.~(1) of the main text.

To match the notation used there, we parameterize the two linear couplings as
\begin{equation}
g=\sqrt{|\alpha_0|^2+|\beta_0|^2},
\qquad
\sin\theta=\frac{\alpha_0}{g},
\qquad
\cos\theta=\frac{\beta_0}{g},
\label{eq:SM_exp_gtheta}
\end{equation}
so that Eq.~\eqref{eq:SM_exp_targetbare} takes the form
\begin{equation}
\hat H_{\rm eff}
=
g\left(\sin\theta\,\hat a^\dagger\hat\sigma^-+\cos\theta\,\hat b^\dagger\hat\sigma^-\right)
+\bar\gamma\,\hat a^\dagger\hat a^\dagger\hat b\,\hat\sigma^-
+\mathrm{H.c.}
\label{eq:SM_exp_targetbare2}
\end{equation}
The quantity directly generated by the circuit-QED scheme is therefore the bare effective three-body coupling $\bar\gamma$. In the main text, however, the nonlinear term is written as $(\gamma/N)\hat a^\dagger\hat a^\dagger\hat b\,\hat\sigma^-$. The corresponding identification is
\begin{equation}
\bar\gamma\equiv \frac{\gamma}{N}.
\label{eq:SM_exp_gamma_ident}
\end{equation}
Thus, the factor $1/N$ is not an additional microscopic suppression mechanism; rather, it is the normalization convention adopted in the sensing Hamiltonian so that $\gamma$ remains the relevant dimensionless nonlinear control parameter when different fixed-$N$ sectors are compared. For a given target excitation sector $N$, one therefore tunes the drive amplitude and longitudinal couplings such that the experimentally realized bare three-body amplitude satisfies Eq.~\eqref{eq:SM_exp_gamma_ident}. With this identification, the circuit-QED implementation and the model analyzed in the main text are fully self-consistent.

In practice, one may first prepare the state $|\downarrow,N,0\rangle$ and keep $\bar\gamma=0$ (equivalently $\gamma=0$) while tuning the ratio of the two linear couplings to the target value of $\theta$. Since the gap is then fixed at $\Delta E=1$, this step does not incur critical slowing down. One may then turn on the drive-induced multiphoton process and ramp $\bar\gamma=\gamma/N$ to enter the desired operating regime. In this way, the experimentally controllable parameters $(g_a,g_b,\Omega,\tilde\omega,\omega_a,\omega_b)$ provide direct access to the effective parameters $(g,\theta,\gamma)$ in the sensing Hamiltonian.

We finally note that the effective Hamiltonian in Eq.~\eqref{eq:SM_exp_targetbare2} is derived in a displaced and driven rotating frame. Accordingly, the local photon-number observables appearing in the effective sensing model should be understood in the same transformed frame. For example, under the displacement transformation in Eq.~\eqref{eq:SM_exp_D}, the bare laboratory-frame photon-number operator $\hat n_a=\hat a^\dagger \hat a$ is mapped to
\begin{equation}
\hat n_a
\ \longleftrightarrow\
\hat D^\dagger \hat n_a \hat D
=
\left(\hat a^\dagger+\frac{\eta_a}{2}\hat\sigma_z\right)
\left(\hat a+\frac{\eta_a}{2}\hat\sigma_z\right),
\label{eq:SM_exp_na_dressed}
\end{equation}
and similarly for $\hat n_b=\hat b^\dagger\hat b$. Since the drive transformation $\hat U_d(t)$ in Eq.~\eqref{eq:SM_exp_Ud} acts only on the qubit degree of freedom and commutes with $\hat\sigma_z$, it does not further modify the transformed local photon-number observable. Therefore, the local photon-number measurement discussed in the main text corresponds, in the laboratory frame, to a dressed local observable involving the same cavity mode and the qubit. In particular, if one chooses to read out cavity $a$ ($b$), the corresponding transformed observable depends only on the displacement parameter $\eta_a$ ($\eta_b$) of that same cavity mode. For implementations in which this displacement parameter is a fixed calibration parameter and remains perturbatively small, the dressed observable differs from the bare cavity photon-number measurement only by perturbatively small corrections. Hence, to leading order, standard photon-number-resolved readout on the chosen single cavity still accesses the effective local observable. If desired, one may also switch off the drive and undo the dressing transformations before readout, thereby mapping the measurement back explicitly to the bare laboratory-frame basis.

\section{Local photon-number measurement saturates the quantum Fisher information in the effective sensing frame}\label{SM5}

In the main text, we stated that, in the effective sensing frame, the QFI of the zero-energy mode can already be saturated by a local photon-number measurement on a single cavity. Here we show this explicitly. The key point is that, because the hopping amplitudes in Eq.~(3) of the main text are all real, the zero-energy-mode amplitudes can be chosen real in the Fock basis.

As shown in Sec.~\ref{SM1}, in the fixed-$N$ sector the zero-energy mode resides entirely on the $\downarrow$ sublattice and can be written as
\begin{equation}
|\psi_0\rangle=\sum_{n=0}^{N} u_n\,|\downarrow,N-n,n\rangle,
\label{eq:SM_zero_mode_measure}
\end{equation}
where the amplitudes $\{u_n\}$ satisfy the recursion relation in Eq.~\eqref{eq:SM_recursion_general}. Since the coefficients $v_n$, $w_n$, and $t_n$ in Eq.~\eqref{eq:SM_hoppings} are all real, this recursion has real coefficients, and the overall phase of the state is arbitrary. Therefore the amplitudes $u_n$ may be chosen real for all $n$.

We now consider a projective measurement of the photon number in cavity $\hat b$. In the zero-energy mode, the outcome $n_b=n$ occurs with probability
\begin{equation}
P_n=|\langle \downarrow,N-n,n|\psi_0\rangle|^2=u_n^2.
\label{eq:SM_Pn_general}
\end{equation}
Because the state has support only on the $\downarrow$ sublattice, measuring $n_b$ on cavity $\hat b$ is equivalent to projecting onto the basis states $|\downarrow,N-n,n\rangle$. Moreover, in the fixed-$N$ sector one may equally view this as a local photon-number measurement on either cavity, since $n_a=N-n_b$.

The corresponding classical Fisher information (CFI) is~\cite{pezze2018quantum}
\begin{equation}
F_\theta^{\rm C}
=
\sum_{n=0}^{N}\frac{(\partial_\theta P_n)^2}{P_n}.
\label{eq:SM_CFI_def}
\end{equation}
Using Eq.~\eqref{eq:SM_Pn_general} and the fact that $u_n$ is real, we have
\begin{equation}
\partial_\theta P_n
=
2u_n\,\partial_\theta u_n,
\label{eq:SM_dPn}
\end{equation}
so that
\begin{equation}
F_\theta^{\rm C}
=
4\sum_{n=0}^{N}(\partial_\theta u_n)^2.
\label{eq:SM_CFI_u}
\end{equation}

On the other hand, for the pure state in Eq.~\eqref{eq:SM_zero_mode_measure}, the QFI is
\begin{equation}
F_\theta^{\rm Q}
=
4\left(
\langle \partial_\theta\psi_0|\partial_\theta\psi_0\rangle
-
|\langle \psi_0|\partial_\theta\psi_0\rangle|^2
\right).
\label{eq:SM_QFI_def_repeat}
\end{equation}
Because the amplitudes are real and the state is normalized,
\begin{equation}
\langle \psi_0|\partial_\theta\psi_0\rangle
=
\sum_{n=0}^{N}u_n\,\partial_\theta u_n
=
\frac{1}{2}\partial_\theta\sum_{n=0}^{N}u_n^2
=
0,
\label{eq:SM_overlap_zero_repeat}
\end{equation}
and therefore
\begin{equation}
F_\theta^{\rm Q}
=
4\sum_{n=0}^{N}(\partial_\theta u_n)^2.
\label{eq:SM_QFI_u}
\end{equation}
Comparing Eqs.~\eqref{eq:SM_CFI_u} and \eqref{eq:SM_QFI_u}, we immediately obtain
\begin{equation}
F_\theta^{\rm C}=F_\theta^{\rm Q}.
\label{eq:SM_CFI_equals_QFI}
\end{equation}

Therefore, in the effective sensing frame, a local photon-number measurement on one cavity already saturates the QFI of the zero-energy mode. In the linear case $\gamma=0$, this result reduces to the binomial distribution discussed in Sec.~\ref{SM1}, while for $\gamma\neq 0$ it continues to hold because the zero-energy-mode amplitudes remain real in the Fock basis. Combined with the dressed-observable discussion in Sec.~\ref{SM4}, this establishes the readout claim made in the main text and clarifies its laboratory-frame interpretation.

\end{document}